\documentclass[12pt]{article}

\usepackage{graphicx}
\usepackage{amsmath}
\usepackage{amssymb}
\usepackage{amsthm}
\usepackage{caption2}
\usepackage{setspace}
\usepackage[natural]{xcolor}

\newtheorem{theorem}{Theorem}[section]

\newtheorem{definition}{Definition}[section]

\DeclareMathOperator*{\doublesum}{\sum\sum}

\def\bi#1#2{ \ba{c} #1 \\ #2 \ea }
\def\mb{\mbox{ }}

\def\ba{\begin{array}}
\def\bd{\begin{document}}
\def\bdes{\begin{description}}
\def\bc{\begin{center}}
\def\be{\begin{equation}}
\def\bea{\begin{eqnarray}}
\def\beaa{\begin{eqnarray*}}
\def\bit{\begin{itemize}}
\def\ben{\begin{enumerate}}
\def\bt{\begin{tabular}}
\def\ea{\end{array}}
\def\ed{\end{document}}
\def\edes{\end{description}}
\def\ec{\end{center}}
\def\ee{\end{equation}}
\def\eea{\end{eqnarray}}
\def\eeaa{\end{eqnarray*}}
\def\et{\end{tabular}}
\def\eit{\end{itemize}}
\def\een{\end{enumerate}}
\def\ub{\underbrace}
\def\ob{\overbrace}
\def\ul{\underline}
\def\bv\begin{verbatim}
\def\ev\end{verbatim}
\def\bal{\begin{align*}}
\def\eal{\end{align*}}

\def\bfv#1{{\bf #1}}
\def\a{\alpha}
\def\b{\beta}
\def\g{\gamma}
\def\d{\delta}
\def\e{\varepsilon}
\def\z{\zeta}
\def\t{\theta}
\def\k{\kappa}
\def\p{\phi}
\def\o{\omega}
\def\G{\Gamma}
\def\D{\Delta}
\def\vp{\varphi}
\def\tdown{\bigtriangledown}
\def\diff{\nabla}
\def\T{\Theta}
\def\L{\Lambda}
\def\S{\Sigma}
\def\P{\Phi}
\def\O{\Omega}
\def\inf{\infty}
\def\prop{\propto}
\def\r|{\right|}
\def\l#1{\left#1}
\def\r#1{\right#1}
\def\ol{\overline}
\def\ltend{\longrightarrow}
\def\tend{\rightarrow}
\def\imply{\Longrightarrow}
\def\X{{\it X }}
\def\Y{{\it Y }}
\def\PB{\Phi (B)}
\def\TB{\Theta (B)}
\def\dd{(1-B)^d}
\def\PBinv{{\Phi (B)}^{-1}}
\def\TBinv{{\Theta (B)}^{-1}}
\def\SST{{\rm SST}}
\def\SSR{{\rm SSR}}
\def\SSE{{\rm SSE}}
\def\MSR{{\rm MSR}}
\def\MST{{\rm MST}}
\def\MSE{{\rm MSE}}
\def\mb{\mbox{ }}
\def\dv{ dependent variable }
\def\iv{ independent variable }
\def\mgf{ moment generating function }
\def\iid{ independent and identically distributed }
\def\pdf{ probability density function }
\def\rv{ random variable }
\def\rvs{ random variables }
\def\varx{{\sigma_x}^2}
\def\vary{{\sigma_y}^2}
\def\rhoxy{{\rho_{x,y}}}
\def\covxy{{Cov( X , Y )}}
\def\bra#1{{( #1_i - \bar #1 )}}
\def\bras#1{{( #1_i - \bar #1 )}^2}
\def\oon{\frac 1{n-1}}
\def\vis#1{#1_i^2}
\def\vbs#1{\bar #1^2}
\def\vsvh#1#2{{\hat #1}_{#2}}
\def\vhi#1{\hat #1_i}
\def\vih#1{\hat #1_i}
\def\bxy#1{( x_#1 , y_#1 )}
\def\tnt{t_{(n-2)}}
\def\tnta{t_{(n-2,\alpha)}}
\def\tntas{t_{(n-2,\alpha /2)}}
\def\veps{\varepsilon}
\def\vepsi{\veps_i}
\def\vepsj{\veps_j}
\def\vareps{\sigma_\veps^2}
\def\var{\sigma^2}
\def\svs#1{s_#1^2}
\def\yih{{\hat y}_i}
\def\binom#1#2{\left(\ba{c} #1 \\ #2 \ea \right)}
\def\bi#1#2{ \ba{c} #1 \\ #2 \ea }
\def\s#1#2{#1_#2}
\def\sxx#1#2#3{#1_{#2 #3}}
\def\sxxx#1#2#3#4{#1_{#2#3#4}}
\def\sxxxx#1#2#3#4#5{#1_{#2#3#4#5}}
\def\sxxxxx#1#2#3#4#5#6{#1_{#2#3#4#5#6}}
\def\sxxxxxx#1#2#3#4#5#6#7{#1_{#2#3#4#5#6#7}}
\def\sxxxxxxx#1#2#3#4#5#6#7#8{#1_{#2#3#4#5#6#7#8}}
\def\h#1{\hat #1}
\def\sh#1#2{{\hat #1}_#2}
\def\sxxh#1#2#3{{\hat #1}_{#2 #3}}
\def\sxxxh#1#2#3#4{{\hat #1}_{#2#3#4}}
\def\sxxxxh#1#2#3#4#5{{\hat #1}_{#2#3#4#5}}
\def\sxxxxxh#1#2#3#4#5#6{{\hat #1}_{#2#3#4#5#6}}
\def\sxxxxxxh#1#2#3#4#5#6#7{{\hat #1}_{#2#3#4#5#6#7}}
\def\sxxxxxxxh#1#2#3#4#5#6#7#8{{\hat #1}_{#2#3#4#5#6#7#8}}
\def\ss#1#2#3{#1_{#2,#3}}
\def\ts#1#2#3{#1_{#2_#3}}
\def\tsxx#1#2#3#4#5{#1_{{#2_#3}#4#5}}
\def\sumv#1#2#3{\sum_{#1=#2}^{{#3}}}
\def\chis{\chi^2}
\def\vsp#1#2#3{#1_#2^#3}
\def\sfss#1#2{S_{#1 + #2}^f}
\def\SS#1#2{{\mbox{SS}}_{#1 #2}}
\def\mb{\mbox{ }}
\def\n0v{N( 0 ,\sigma^2 )}
\def\nmv{N( \mu ,\sigma^2 )}
\def\bvsv#1#2{\{ #1_#2 \}}
\begin{document}

\thispagestyle{empty}
\baselineskip=26pt
\begin{center}
{\bf \Large {\bf A New Test of Multivariate Nonlinear Causality }}\large
\\
\renewcommand{\thefootnote}{\fnsymbol{footnote}}
\vskip 1ex
{\bf Zhidong Bai} \\  KLASMOE and School of Mathematics and Statistics, \\ Northeast Normal University, China \\
\vskip 1ex {\bf Yongchang Hui}\footnote{Corresponding author: Yongchang Hui, School of Mathematics and Statistics,
Xi'an Jiaotong University. No.28, Xianning West Road, Xi'an, Shaanxi, P.R. China.
Tel: (86)-029-82663170, Email: huiyc180@xjtu.edu.cn} \\ School of Mathematics and Statistics, \\ Xi'an Jiaotong University, China \\
\vskip 1ex {\bf Zhihui Lv} \\   KLASMOE and  School of Mathematics and Statistics, \\ Northeast Normal University, China \\
\vskip 1ex {\bf Wing-Keung Wong} \\ Department of Finance, Asia University, Taiwan \\
Department of Economics, Lingnan University, Hong Kong \\
\vskip 1ex {\bf Shurong Zheng} \\   KLASMOE and  School of Mathematics and Statistics, \\ Northeast Normal University, China \\
\vskip 1ex {\bf Zhen-Zhen Zhu} \\   KLASMOE and  School of Mathematics and Statistics, \\ Northeast Normal University, China \\
\end{center}

\newpage

\newpage \setcounter{page}{1}
\baselineskip = 20pt

\begin{center}
{\bf \Large {\bf A New Test of Multivariate Nonlinear Causality}}
\end{center}

\vspace{0.2in} \noindent \bf Abstract \rm $\quad$
The multivariate nonlinear Granger causality developed by Bai et al. (2010) plays an important
role in detecting the dynamic interrelationships between two groups of variables. Following the idea of Hiemstra-Jones (HJ) test
proposed by Hiemstra and Jones (1994), they attempt to establish a central limit theorem (CLT) of their test statistic by applying the asymptotical property of multivariate $U$-statistic. However, Bai et al. (2016) revisit the HJ test and find that the test statistic given by HJ is NOT a function of $U$-statistics which implies that the CLT neither proposed by Hiemstra and Jones (1994) nor the one extended by Bai et al. (2010) is valid for statistical inference. In this paper, we re-estimate the probabilities and reestablish the CLT of the new test statistic. Numerical simulation shows that our new estimates are consistent and our new test performs decent size and power.

\vspace{0.2in} \noindent \bf Keywords: \rm $\quad$ nonlinear Granger causality, Hiemstra-Jones test, multivariate


\thispagestyle{empty}

\newpage

\section{Introduction}\label{introduction}
\setcounter{footnote}{0}
After the pioneering work of Granger (1969), Granger causality tests have been developed into a set of useful methods to detect causal relations between time series in economics and finance. Linear Granger causality tests within the linear autoregressive
model class have been developed in many directions, e.g., Hurlin et al. (2001) proposed
a procedure for causality tests with panel data, Ghysels et al. (2016)
test for Granger causality with mixed frequency data based on the multiple-horizon framework
established by Dufour and Renault (1998) and Dufour et al. (2006).
Though linear tests of Granger causality have been investigated very deeply, they are limited in their capability to
detect nonlinear causality.

The real world is ``almost certainly nonlinear'' as Granger (1989) notes, so it is more important to test
the nonlinear causality. Baek and Brock (1992) develop a nonlinear Granger causality test which is modified by Hiemstra and Jones (1994) later on to study the bivariate nonlinear causal relationship between two series.
Among the various tests of nonlinear Granger causality, the Hiemstra-Jones test (hereafter, the HJ test) proposed by Hiemstra and Jones (1994)
is the most cited by scholars and the most frequently applied by practitioners in economics and finance.
There were over 1100 Google Scholar hits by September 2016, which illustrates its significance in the economics and finance literatures.

Bai et al. (2010) extend the HJ test from bivariate setting to multivariate setting catering to the practical needs that economic and financial factors
usually move together and influence others in groups. This extension encourages a large amount of applications. For example, Lam et al. (2012) suggest to use such technics to make better investment decisions. Zheng and Chen (2013) proposed a complete double selection method in identifying external influential factors for a particular stock market. Choudhry et al. (2015) investigates the nonlinear dynamic co-movements between gold returns, stock market returns and stock market volatility during the recent global financial crisis. Choudhry et al. (2016) investigate the relationship between stock market volatility and the business cycle in four major economies US, Canada, Japan and the UK.

However, several works note that counterintuitive results are obtained from the HJ test, Diks and Panchenko (2005, 2006) find that the HJ test is seriously over-rejecting in simulation studies. In accordance with the evidence presented by Diks and Panchenko (2005, 2006), Bai et al. (2016) reinvestigate the HJ test and reveal some of the underlying reasons for the questionable performance of HJ test. They find that the estimators of the probabilities in the definition are not $U$-statistics as Hiemstra and Jones (1994) claimed and the central limit theorem of the test statistics is not valid. Bai et al. (2016) propose a set of consistent estimators of the probabilities in the definition of Hiemstra and Jones (1994) and provide a new test statistic with its asymptotic distribution.

Considering the significant importance of the multivariate nonlinear Granger causality test, there is an urgent need to reinvestigate Bai et al. (2010)
 and extend Bai et al. (2016) to multivariate setting. The remainder of this paper is organized as follows. In Section 2, we simply review the procedure of the multivariate nonlinear Granger causality test (here after BWZ test) extended by Bai et al. (2010). In Section 3, we re-estimate the probabilities in the definition of Bai et al. (2010) and establish the asymptotic distribution of the new test statistics. Simulation results are presented in Section 4. Finally, we provide some concluding remarks in Section 5.

\vspace{.2in}

\section{The Multivariate Nonlinear Causality Test Extended from HJ Test}\label{BivariateNLCT}
\bigskip
Bai et al. (2010) consider two strictly stationary and weakly dependent vector time series processes
${X}_{t} = ({X}_{1,t},{X}_{2,t},\cdots,{X}_{n_1,t})'$, ${Y}_{t} = ({Y}_{1,t},{Y}_{2,t},\cdots,{Y}_{n_2,t})'$.
The $m_{x_{i}}$-length lead vector of $X_{i,t}$ is defined as ${X}^{m_{x_{i}}}_{i,t}\equiv (X_{i,t},X_{i,t+1},\cdots,X_{i,t+m_{x_{i}}-1}), m_{x_{i}}=1,2,\cdots, t=1,2, \cdots , \\$ similarly $L_{x_{i}}$-length lag vector of $X_{i,t}$ , and $L_{y_{i}}$-length lag vector of $Y_{i,t}$ are defined as
${X}^{L_{x_{i}}}_{i,t-L_{x_{i}}}\equiv(X_{i,t-L_{x_{i}}},X_{i,t-L_{x_{i}}+1},\cdots,X_{i,t-1}),L_{x_{i}}=1,2,\cdots,t=L_{x_{i}}+1,L_{x_{i}}+2,
\cdots ,\\$
${Y}^{L_{y_{i}}}_{i,t-L_{y_{i}}}\equiv(Y_{i,t-L_{y_{i}}},Y_{i,t-L_{y_{i}}+1},\cdots,Y_{i,t-1}), L_{y_{i}}=1,2,\cdots, t=L_{y_{i}}+1,L_{y_{i}}+2,
\cdots$ .\\
Denote $M_{x}=( m_{x_{1}},\cdots, m_{x_{n_{1}}})$, $m_{x}=\max(m_{x_{1}},\cdots,m_{x_{n_{1}}})$, $L_{x}=(L_{x_{1}},\cdots,L_{x_{n_{1}}})$,  $l_{x}=\max(L_{x_{1}},\cdots,L_{x_{n_{1}}})$, $L_{y}=(L_{y_{1}},\cdots,L_{y_{n_{2}}})$,  $l_{y}=\max(L_{y_{1}},\cdots,L_{y_{n_{2}}})$. For given $M_{x},L_{x},L_{y},e$, Bai et al. (2010) define that
\begin{align*}
\{\|{X}^{M_{x}}_t-{X}^{M_{x}}_s\|<e \}\equiv&\{ \|{X}^{m_{x_{i}}}_{i,t}-{X}^{m_{x_{i}}}_{i,s}\|<e,\quad for\quad all \quad i=1,2,\cdots,n_{1}\}\\
\{\|{X}^{L_{x}}_{t-L_{x}}-{X}^{L_{x}}_{s-L_{x}}\|<e \}\equiv&\{ \|{X}^{L_{x_{i}}}_{i,t-L_{x_{i}}}-{X}^{L_{x_{i}}}_{i,s-L_{x_{i}}}\|<e,\quad for\quad all \quad i=1,2,\cdots,n_{1}\}\\
\{\|{Y}^{L_{y}}_{t-L_{y}}-{Y}^{L_{y}}_{s-L_{y}}\|<e\} \equiv&\{ \|{Y}^{L_{y_{i}}}_{i,t-L_{y_{i}}}-{Y}^{L_{y_{i}}}_{i,s-L_{y_{i}}}\|<e,\quad for\quad all \quad i=1,2,\cdots,n_{2}\} \, ,
\end{align*}
where $\|\cdot\| $ denotes the maximum norm
defined as $\|X-Y\|= \max  \big ( |x_1-y_1|, |x_2-y_2|, \cdots,
|x_n-y_n| \big )$ for any two vectors $X= \big (x_1, \cdots, x_n
\big )$ and $Y= \big (y_1, \cdots, y_n \big )$.

\begin{definition}  \label{defsinglenonlinear} \quad
The vector time series $\{Y_{t}\}$ does not strictly Granger cause another vector time series $\{X_{t}\}$ if
\begin{align} &P\left( \parallel{X}^{M_{x}}_t-{X}^{M_{x}}_s \parallel<e
\big | \parallel{X}^{L_{x}}_{t-L_{x}}-{X}^{L_{x}}_{s-L_{x}}\parallel<e,\parallel{Y}^{L_{y}}_{t-L_{y}}-{Y}^{L_{y}}_{s-L_{y}}\parallel<e \right) \nonumber \\
& = P\left( \parallel{X}^{M_{x}}_t-{X}^{M_{x}}_s \parallel<e
\big | \parallel{X}^{L_{x}}_{t-L_{x}}-{X}^{L_{x}}_{s-L_{x}} \parallel<e \right) \, , \label{test0}
\end{align}
where $P(\cdot|\cdot)$ denotes conditional probability.
\end{definition}

Using the notation
\begin{align*}
{C}_1 (M_{x}+L_{x},L_{y},e)
&\equiv P\left(\parallel{X}^{M_{x}+L_{x}}_{t-L_{x}}-{X}^{M_{x}+L_{x}}_{s-L_{x}}\parallel<e,\parallel{Y}^{L_{y}}_{t-L_{y}}-{Y}^{L_{y}}_{s-L_{y}}\parallel<e\right)\\
{C}_2 (L_{x},L_{y},e)&\equiv P\left(\parallel{X}^{L_{x}}_{t-L_{x}}-{X}^{L_{x}}_{s-L_{x}}\parallel<e
,\parallel{Y}^{L_{y}}_{t-L_{y}}-{Y}^{L_{y}}_{s-L_{y}}\parallel<e\right)\\
{C}_3 (M_{x}+L_{x},e)&\equiv P\left(\parallel{X}^{M_{x}+L_{x}}_{t-L_{x}}-{X}^{M_{x}+L_{x}}_{s-L_{x}}\parallel<e\right)\\
{C}_4 (L_{x},e)& \equiv P\left(\parallel{X}^{L_{x}}_{t-L_{x}}-{X}^{L_{x}}_{s-L_{x}}\parallel<e\right) \, ,
\end{align*}
Bai, et al. (2010) re-express Equation (\ref{test0}) as
\begin{equation}
\frac{{C}_1 (M_{x}+L_{x},L_{y},e)}{{C}_2 (L_{x},L_{y},e)} = \frac{{C}_3 (M_{x}+L_{x},e)}{{C}_4 (L_{x},e)} \, .
\end{equation}

For two sets of simultaneous samples $\{x_{i,t},i=1,\cdots,n_{1}, t=1, \cdots, T\}$ and $\{y_{i,t},i=1,\cdots,n_{2}, t=1, \cdots, T\}$, they propose the following test statistic
\begin{equation}
\sqrt{n}\left(\frac{C_1 \big (M_{x}+{L_x},L_y,e,n \big )}{C_2 \big
({L_x},L_y,e,n
 \big )}-\frac{C_3 \big (M_{x}+{L_x},e,n \big )}{C_4 \big ({L_x},e,n \big )}\right)\, , \label{test1}
\end{equation}
where
\begin{align*}
& C_1  \big (M_{x}+{L_x},L_y,e,n  \big )\\
& \equiv
\frac{2}{n(n-1)}\doublesum_{t<s} \prod^{n_{1}}_{i=1} I \left({x}^{m_{x_{i}}+L_{x_{i}}}_{i,t-L_{x_{i}}},{x}^{m_{x_{i}}+L_{x_{i}}}_{i,s-L_{x_{i}}},e \right) \cdot \prod^{n_{2}}_{i=1} I \left({y}^{L_{y_{i}}}_{i,t-L_{y_{i}}},{y}^{L_{y_{i}}}_{i,s-L_{y_{i}}},e \right) , \\
& C_2  \big ({L_x},L_y,e,n  \big ) \equiv
\frac{2}{n(n-1)}\doublesum_{t<s} \prod^{n_{1}}_{i=1} I \left({x}^{L_{x_{i}}}_{i,t-L_{x_{i}}},{x}^{L_{x_{i}}}_{i,s-L_{x_{i}}},e \right) \cdot \prod^{n_{2}}_{i=1} I \left({y}^{L_{y_{i}}}_{i,t-L_{y_{i}}},{y}^{L_{y_{i}}}_{i,s-L_{y_{i}}},e\right),\\
& C_3 \big (M_{x}+{L_x},e,n \big )\equiv
\frac{2}{n(n-1)}\doublesum_{t<s}
 \prod^{n_{1}}_{i=1} I \left({x}^{m_{x_{i}}+L_{x_{i}}}_{i,t-L_{x_{i}}},{x}^{m_{x_{i}}+L_{x_{i}}}_{i,s-L_{x_{i}}},e \right) ,\\
& C_4 \big ({L_x},e,n \big )\equiv \frac{2}{n(n-1)}\doublesum_{t<s}
 \prod^{n_{1}}_{i=1} I \left({x}^{L_{x_{i}}}_{i,t-L_{x_{i}}},{x}^{L_{x_{i}}}_{i,s-L_{x_{i}}},e \right) ,\\
\, \mbox{and} \\ & I(x,y,e)=
\begin{cases} 0, & \text{if $\|x-y\|>e$}\\
1, & \text{if $\|x-y\|\leq e$}
\end{cases} \, .
\end{align*}
\textbf{Remark:} Following the instruction of Hiemstra and Jones (1994), Bai et al. (2010) take $C_j(*,n)$s as multivariate $U$-statistic estimators of their counterparts $C_j(*)$s and apply the asymptotic property of $U$-statistic to show the limiting results for the test statistics (\ref{test1}).
However the $C_j(*,n)$s are not $U$-statistics, because the expectations of the general terms are not the same. Moreover, the
$C_j(*)$s are related to the indices $t$ and $s$ (in fact, related to $|t-s|$ for strongly stationary processes), while the $C_j(*,n)$s were independent of $t$ and $s$ for summing up over them. 
 Therefore, the $C_j(*,n)$ estimators are neither consistent nor asymptotic normal estimators of their counterparts $C_j(*)$.

\section{ A New Multivariate Nonlinear Causality Test}\label{BivariateNLCT}
We first remind the reader that the pair $(s,t)$ (in fact, $|t-s|$ for strongly stationary processes) in Equation (\ref{test0})
of Definition \ref{defsinglenonlinear} is a key parameter of the probabilities $C_j(*)$.
In fact, both   Hiemstra and Jones (1994) and Bai et al. (2010) note this, and there is no problem in Equation (\ref{test0})
of Definition \ref{defsinglenonlinear}. However, it seems that they overlooked this fact in their proposed estimation of $C_j(*)$. The improper estimators $C_j(*, n)$ thus lead to an invalid asymptotic distribution of the test statistic.

We now begin to state the procedure for our new test. For any given pair $(s,t)$, we denote
\begin{align*}
& {C}_1 (M_{x}+L_{x},L_{y},e;t,s)\equiv P\left(\parallel{ X}^{M_{x}+L_{x}}_{t-L_{x}}-{X}^{M_{x}+L_{x}}_{s-L_{x}}\parallel<e,\parallel{Y}^{L_{y}}_{t-L_{y}}-{ Y}^{L_{y}}_{s-L_{y}}\parallel<e\right)\\
& {C}_2 (L_{x},L_{y},e;t,s)\equiv P\left(\parallel{ X}^{L_{x}}_{t-L_{x}}-{ X}^{L_{x}}_{s-L_{x}}\parallel<e
,\parallel{ Y}^{L_{y}}_{t-L_{y}}-{ Y}^{L_{y}}_{s-L_{y}}\parallel<e\right)\\
& { C}_3 (M_{x}+L_{x},e;t,s)\equiv P\left(\parallel{ X}^{M_{x}+L_{x}}_{t-L_{x}}-{ X}^{M_{x}+L_{x}}_{s-L_{x}}\parallel<e\right)\\
& { C}_4 (L_{x},e;t,s) \equiv P\left(\parallel{ X}^{L_{x}}_{t-L_{x}}-{ X}^{L_{x}}_{s-L_{x}}\parallel<e\right)
\end{align*}
Under the assumption of the stationary, for the given pair $(s,t)$, if $s-t=l$, we denote
${C}_1 (M_{x}+L_{x},L_{y},e;t,s)\equiv {C}_1 (M_{x}+L_{x},L_{y},e;t,l)$, which does not depend on t, so we can write ${C}_1 (M_{x}+L_{x},L_{y},e;l)$ instead of ${C}_1 (M_{x}+L_{x},L_{y},e,t;l)$, the same to the others.
So under the assumption of strictly stationary, for each $l>0$,
we examine whether there is nonlinear Granger causality from $\{Y_t\}$ to $\{X_t\}$ by testing the following hypothesis
\begin{equation}\label{newh0}
H_{0} \  : \ \ \frac{{C}_1 (M_{x}+L_{x},L_{y},e;l)}{{C}_2 (L_{x},L_{y},e;l)}=\frac{{C}_3 (M_{x}+L_{x},e;l)}{{C}_4(L_{x},e;l)} \, .
\end{equation}

If we consider two sets of simultaneous samples\footnote{To implement the test, each series is standardized so that all series share a common
standard deviation, and thereby share a common scale parameter.} $\{x_{i,t},i=1,\cdots,n_{1}, t=1, \cdots, T\}$ and $\{y_{j,t},j=1,\cdots,n_{2}, t=1, \cdots, T\}$,
we first provide the consistent estimators of ${C}_1 (M_{x}+L_{x},L_{y},e;l)$, ${C}_2 (L_{x},L_{y},e;l)$, ${C}_3 (M_{x}+L_{x},e;l)$ and ${C}_4 (L_{x},e;l)$ are
\begin{align*}
& {\hat{C}}_1 (M_{x}+L_{x},L_{y},e;l)\equiv \frac{1}{n}\sum^{T-l-m_{x}+1}_{t=L_{xy}+1}\prod^{n_{1}}_{i=1}I\left({x}^{m_{x_{i}}+L_{x_{i}}}_{i,t-L_{x_{i}}},{x}^{m_{x_{i}}+L_{x_{i}}}_{i,t+l-L_{x_{i}}},e\right)\cdot\prod^{ n_{2}}_{j=1}I\left({y}^{L_{y_{j}}}_{j,t-L_{y_{j}}},{y}^{L_{y_{j}}}_{j,t+l-L_{y_{j}}},e\right) \\
& {\hat{C}}_2 (L_{x},L_{y},e;l)\equiv \frac{1}{n}\sum^{T-l-m_{x}+1}_{t=L_{xy}+1}\prod^{n_{1}}_{i=1}I\left({ x}^{L_{x_{i}}}_{i,t-L_{x_{i}}},{x}^{L_{x_{i}}}_{i,t+l-L_{x_{i}}},e\right)\cdot\prod^{n_{2}}_{j=1}I\left({ y}^{L_{y_{j}}}_{j,t-L_{y_{j}}},{y}^{L_{y_{j}}}_{j,t+l-L_{y_{j}}},e\right)\\
& {\hat{C}}_3 (M_{x}+L_{y},e;l)\equiv \frac{1}{n}\sum^{T-l-m_{x}+1}_{t=L_{xy}+1}\prod^{n_{1}}_{i=1}I\left({ x}^{m_{x_{i}}+L_{x_{i}}}_{i,t-L_{x_{i}}},{x}^{m_{x_{i}}+L_{x_{i}}}_{i,t+l-L_{x_{i}}},e\right)\\
& {\hat{C}}_4 (L_{x},e;l)\equiv \frac{1}{n}\sum^{T-l-m_{x}+1}_{t=L_{xy}+1}\prod^{n_{1}}_{i=1}I\left({ x}^{L_{x_{i}}}_{i,t-L_{x_{i}}},{x}^{L_{x_{i}}}_{i,t+l-L_{x_{i}}},e\right)\\
\end{align*}
where $ L_{xy} = \max(L_{x},L_{y})$, $n=T- L_{xy}-l-m_{x}+1$ and
$$I(x,y,e)\equiv
\begin{cases}
0, & \text{if $\|x-y\|>e$ }\\
1, & \text{if $\|x-y\|\leq e$ }
\end{cases} .$$

The consistency of our proposed estimators can be shown straightforwardly and the detail of the proof is omitted.
We use a simple numerical study to show that our estimators are consistent whereas those of Bai et al. (2010) are not.
Let$$ \left(
                                                                                                                        \begin{array}{c}
                                                                                                                          X_{1,t} \\
                                                                                                                          X_{2,t} \\
                                                                                                                        \end{array}
                                                                                                                      \right)=\left(
                                                                                                                                \begin{array}{cc}
                                                                                                                                   1& 0 \\
                                                                                                                                  0 & 1 \\
                                                                                                                                \end{array}
                                                                                                                              \right)\left(
                                                                                                                                        \begin{array}{c}
                                                                                                                                          a_{1,t-1} \\
                                                                                                                                          a_{2,t-1}\\
                                                                                                                                        \end{array}
                                                                                                                                      \right) +\left(
   \begin{array}{c}
     a_{1,t} \\
     a_{2,t}\\
   \end{array}
 \right) ,$$ where $ \{a_{1,t}\}, \{a_{2,t}\} $ are i.i.d. and mutually independent random variables generated from $ N(0,1)$, while \{$Y_t$\} could be any stationary sequence. Let $l = 1$, $L_x = L_y = M_x = 1$. We can calculate the exact values of ${C_4 \big ({L_x},e;l  \big)}$, which are $0.2709$ and $0.5057$, respectively, when $e= 1$ and $e= 1.5$. For simplicity, we denote the true value of ${C_4 \big ({L_x},e;l  \big)}$, the estimate proposed by Bai et al.(2010) and our new estimate as $C_4$, $\hat{C}_{4}^{BWZ}$ and $\hat{C}_4$, respectively, in Table 1. Additionally, Table 1 provides the estimated values with their corresponding relative estimation errors in brackets when $T= 1000, 2000 \, \textrm{and} \, 4000$. It is obvious that $\hat{C}_{4}^{BWZ}$ is not consistent.

\begin{table}[!htbp]\scriptsize
\begin{center}
\caption{${C_4 \big ({L_x},e;l  \big)}$ and estimated values.}  \label{table5} \vskip 0.1in
\begin{tabular}{|c|ccccccccc|}
\hline \multicolumn{1}{|c|}{} &\multicolumn{1}{c}{}
&\multicolumn{3}{c}{$e=1$} &\multicolumn{1}{c}{}
&\multicolumn{3}{c}{$e=1.5$} &\multicolumn{1}{c|}{}\\
\cline{3-5}\cline{7-9}
$T =$ && $C_4$ &$\hat{C}_4$&$\hat{C}_{4}^{BWZ}$&&$C_4$&$\hat{C}_4$& $\hat{C}_{4}^{BWZ}$& \\
\hline
1000&&0.2709 &0.2497(7.83\%)&0.0154(94.32\%)&&0.5057 &0.4774(5.60\%) & 0.0732(85.53\%)&\\
\hline
2000&&0.2709 &0.2639(2.58\%)&0.0176(93.50\%)&&0.5057 &0.4847(4.15\%) & 0.0795(84.28\%)& \\
\hline
4000&&0.2709 &0.2692(0.63\%)&0.0193(92.88\%)&&0.5057 &0.4909(2.93\%) & 0.0820 (83.78\%)&\\
\hline
\end{tabular}
\begin{flushleft}
{\small Note:
The true value of ${C_4 \big ({L_x},e;l  \big)}$ is denoted $C_4$, the BWZ estimate and our new estimate are denoted $\hat{C}_{4}^{BWZ}$ and $\hat{C}_4$, respectively. The relative estimation errors are in the accompanying brackets.}
\end{flushleft}
\end{center}
\end{table}

\bigskip
Now, we propose
\begin{equation} \label{newtest}
T_n = \sqrt{n}\left(\frac{\hat{C}_1 \big ({M_x}+{L_x},L_y,e,l \big )}{\hat{C}_2 \big
({L_x},L_y,e,l \big )}-\frac{\hat{C}_3 \big ({M_x}+{L_x},e,l \big )}{\hat{C}_4 \big
({L_x},e,l \big )}\right)
\end{equation}
as the test statistic, and we derive the following asymptotic distribution of $T_n$ for the Granger causality test.

\begin{theorem} \label{newBHWtest} \quad Stationary sequences  $\{x_{i,t},i=1,\cdots,n_{1}, t=1, \cdots, T\}$ and $\{y_{j,t},j=1,\cdots,n_{2}, t=1, \cdots, T\}$ are strong mixing, with mixing coefficients satisfying the conditions of Lemma 1 presented in Appendix, for given values of
$l, L_x, L_y, M_x$ and $e>0$, under the null hypothesis that $\{Y_t\}$ does not strictly
Granger cause $\{X_t\}$, then the test statistic is defined in (\ref{newtest})
\begin{equation*}
\sqrt{n}\left(\frac{\hat{C}_1 \big ({M_x}+{L_x},L_y,e,l \big )}{\hat{C}_2 \big
({L_x},L_y,e,l \big )}-\frac{\hat{C}_3 \big ({M_x}+{L_x},e,l \big )}{\hat{C}_4 \big
({L_x},e,l \big )}\right) \overset{d}{\longrightarrow} N \big(0, \sigma^2
(M_x,{L_x},L_y,e,l) \big) \, .
\end{equation*}
\end{theorem}

The asymptotic variance $\sigma^2(M_x,{L_x},L_y,e,l)$ with its consistent
estimator $\hat{\sigma}^2(M_x,{L_x},L_y,e,l)$ and the proof of theorem \ref{newBHWtest} are given in the Appendix. The hypothesis $H_0$ defined in (\ref{newh0}) is rejected at $\alpha$ if
\begin{eqnarray*}
\big| T_n \big |/{\hat{\sigma}(M_x,{L_x},L_y,e,l)} > z_{{\alpha/2}},
\end{eqnarray*}
where $z_{{\alpha/2}}$ is the up ${\alpha/2}$ quantile of the standard normal distribution.
In this situation, we will conclude that there exists nonlinear Granger causality from \{$Y_{t}$\} to \{$X_{t}$\}.

There are several possible methods to estimate the asymptotic covariance $\sigma^2(M_x,{L_x},L_y,e,l)$. A model-based approach uses known laws of \{$X_t$\} and \{$Y_t$\} to calculate the expectations in the formula given in the Appendix and simply substitutes $C_j(*), j = 1, 2, 3,4$ with their corresponding estimates. However, in practice, we can hardly avoid model misspecification and may obtain improper laws of \{$X_t$\} and \{$Y_t$\}. We suggest the use of bootstrap methods as in the simulation studies we use to test hypothesis $H_0$.

\section{Simulation}
In this section, we perform numerical studies using simulations to illustrate the applicability
and superiority of the new multivariate nonlinear Granger causality test developed in Section 3.
Let $R$ be the times of rejecting the null hypothesis that $Y_t$ does not strictly Granger cause $X_t$ nonlinearly in 10,000 replications at the
$\alpha$ level, and thus, the empirical power is ${R}/{10,000}$. In our simulation, the level $\alpha=0.05$, we standardized the series and chose the same lag length and lead length: $L_x = L_y = M_x = 1$. We set three situations of $l$ and two situations of $e$: $l=1$, $l=2$, $l=3$ and $e=1$, $e=1.5$.

Consider the following model:
\begin{align}\label{testmodel}
X_{t}=\beta Y_{1,t-1}Y_{2,t-1}+\varepsilon_{t} \, ,
\end{align}
where $\{(Y_{1,t},Y_{2,t})^{\prime}\}$ are i.i.d. and mutually independent random variables generated from standard normal distribution $N(0,1)$, $\{\varepsilon_{t}\}$ is
Gaussian white noise generated from $N(0, 0.1)$ and independent of  $\{Y_{1,t}\}, \{Y_{2,t}\}$.
There is no nonlinear Granger causality from $Y_t$ to $X_t$ when $\beta= 0$, and causality strengthens when $\beta$
increases.

From the results displayed in Table 2, we conclude first that our test possesses decent size, as we can see when $\beta = 0$ the empirical size are all closed to the test level 0.05 for different settings of parameters and sample size. Second, our test possesses very appropriate power, as we see that empirical power increases as $\beta$ increases, especially when sample size is 500 the empirical power sharply increase to 1. Further, we find that different settings of $e$ may influence the test results. Though the influence is little in our simulation, we still suggest that practitioners choose a couple of different values of $e$.

\begin{table}\label{simure}
\centering
\caption{Test multivariate nonlinear Granger causality form $Y_t$ to $X_t$} \vskip 0.1in
\begin{tabular}{|c|ccccccccc|}
\hline \multicolumn{1}{|c|}{$T = 200$} &\multicolumn{1}{c}{}
&\multicolumn{3}{c}{$e=1$} &\multicolumn{1}{c}{}
&\multicolumn{3}{c}{$e=1.5$} &\multicolumn{1}{c|}{}\\
\cline{3-5}\cline{7-9}
 && $l=1$ &$l=2$&$l=3$&&$l=1$&$l=2$& $l=3$& \\
\hline
$\beta=0 \ \ $ && 0.0419 & 0.0441 & 0.0432 && 0.0444 & 0.0506 & 0.0438 &\\
$\beta=0.1$ && 0.1509 & 0.2341 & 0.2184 && 0.3647 & 0.5209 & 0.5121 &\\
$\beta=0.2$ && 0.5629 & 0.7425 & 0.7284 && 0.8352 & 0.9416 & 0.9416 &\\
$\beta=0.3$ && 0.8178 & 0.9323 & 0.9267 && 0.929 & 0.9810 & 0.9825 &\\
$\beta=0.4$ && 0.8994 & 0.9712 & 0.9719 && 0.9512 & 0.9878 & 0.9889 &\\
$\beta=0.5$ && 0.9366 & 0.9812 & 0.9808 && 0.9586 & 0.9914 & 0.9915 &\\
$\beta=0.6$ && 0.9475 & 0.9870 & 0.9862 && 0.9640 & 0.9918 & 0.9940 &\\
$\beta=0.7$ && 0.9574 & 0.9875 & 0.9874 && 0.9664 & 0.9921 & 0.9942 &\\
$\beta=0.8$ && 0.9615 & 0.9888 & 0.9882 && 0.9688 & 0.9926 & 0.9953 &\\
$\beta=0.9$ && 0.9633 & 0.9896 & 0.9901 && 0.9711 & 0.9923 & 0.994 &\\
\hline
\hline \multicolumn{1}{|c|}{$T = 500$} &\multicolumn{1}{c}{}
&\multicolumn{3}{c}{$e=1$} &\multicolumn{1}{c}{}
&\multicolumn{3}{c}{$e=1.5$} &\multicolumn{1}{c|}{}\\
\cline{3-5}\cline{7-9}
 && $l=1$ &$l=2$&$l=3$&&$l=1$&$l=2$& $l=3$& \\
\hline
$\beta=0 \ \ $ && 0.0560 & 0.0560 & 0.0478 && 0.0501 & 0.0529 & 0.0423 &\\
$\beta=0.1$ && 0.3969 & 0.5081 & 0.5344 && 0.7835 & 0.9043 & 0.9017 &\\
$\beta=0.2$ && 0.9622 & 0.9908 & 0.9900 && 0.9994 & 1.0000 & 1.0000 &\\
$\beta=0.3$ && 0.9989 & 1.0000 & 0.9997 && 0.9999 & 1.0000 & 1.0000 &\\
$\beta=0.4$ && 0.9998 & 1.0000 & 1.0000 && 1.0000 & 1.0000 & 1.0000 &\\
$\beta=0.5$ && 1.0000 & 1.0000 & 1.0000 && 1.0000 & 1.0000 & 1.0000 &\\
$\beta=0.6$ && 1.0000 & 1.0000 & 1.0000 && 1.0000 & 1.0000 & 1.0000 &\\
$\beta=0.7$ && 1.0000 & 1.0000 & 1.0000 && 1.0000 & 1.0000 & 1.0000 &\\
$\beta=0.8$ && 1.0000 & 1.0000 & 1.0000 && 1.0000 & 1.0000 & 1.0000 &\\
$\beta=0.9$ && 1.0000 & 1.0000 & 1.0000 && 1.0000 & 1.0000 & 1.0000 &\\
\hline
\end{tabular}
\begin{flushleft}
\small{Note: $Y_t$ and $X_t$ are from the model present in equation (\ref{testmodel}). $L_x = L_y = M_x = 1$ in our test. Simulation is conducted with the test level $\alpha= 5\%$, and 10,000
replications.}
\end{flushleft}
\end{table}

\newpage
\vspace{.2in}\section{Conclusion and Remarks}\label{conclusion}
In this paper, we reinvestigate the multivariate nonlinear Granger causality test extended by Bai et al. (2010) which attempt to uncover significant nonlinearities in the dynamic interrelationships between two groups of variables. We find that Bai et al. (2010) as well as Hiemstra and Jones (1994) take the estimators of the probabilities in their definition as $U$-statistics and establish a CLT of the test statistic by applying the asymptotic property of $U$-statistics. After revealing that the estimators proposed by Bai et al. (2010) is not $U$- statistics, we show that their estimators are also not consistent.

The procedure of our new test begins with presenting consistent estimators of probabilities in the definition. Numerical study supports that our estimators are consistent, further our new test possesses admirable properties both in size and power.

There are still amounts of appealing aspects in nonlinear Granger causality test. It is worth noting that Diks and Wolski (2015) extend the test in Diks and Panchenko (2006) which highlight a need for substitutions for the relationship tested in the HJ test.

\section*{Appendix}

\setcounter{equation}{0}

\def\theequation{A.\arabic{equation}}

\bigskip
\noindent

\bigskip \noindent
{\large \bf A1: Central Limit Theorems for strong mixing stationary sequence}

\bigskip
\noindent
$\{\left(Z_t, \mathcal{F}_t\right), -\infty < t < \infty\}$ is a stochastic process defined on the probability space $(\Omega,\mathcal{F},P)$.
The history and the future of $Z_t$  are $\sigma$-algebras $\mathfrak{M}_{t}^\infty = \{\mathcal{F}_{s},s > t\}$ and $\sigma$-algebras
$\mathfrak{M}_{-\infty}^{t} = \{\mathcal{F}_{s},s < t\}$ respectively.

Let \{$\left(Z_i, \mathcal{F}_i\right)$\} be a stationary sequence with $E(Z_i)=0$, $E({Z_i}^2)<0$,and set $S_n^m=\sum\limits_{i=m+1}^{n+m}Z_i$, ${\sigma_n}^2=Var(S_n^m)$.We shall say that the sequence satisfies the central limit theorem if
\begin{eqnarray*}
\lim_{n\to\infty}P\{\frac{S_n^m}{\sigma_n}<z\}=(2\pi)^{-\frac{1}{2}}\int_{-\infty}^{z}{e^{-\frac{1}{2}u^2}du}=\Phi(z) \ .
\end{eqnarray*}

\bigskip \noindent
{\bf Definition A1:}{ A stationary process \{$Z_t$\} is said to be strongly mixing (completely regular) if
$\alpha(\tau)=\sup\limits_{A\in {\mathfrak{m}_{-\infty}^0}, {B \in \mathfrak{m}_\tau^\infty}}|P(AB)-P(A)P(B)|\to 0$
as $\tau \to \infty$ through positive values.}

\noindent \textbf{Lemma A1}: Let the stationary sequence \{$Z_i$\} satisfy the strong mixing condition with mixing coefficient $\alpha(n)$, and let $E|Z_i|^{2+\delta}<\infty$ for some $\delta>0$. If $\sum\limits_{n=1}^{\infty}{\alpha(n)}^{\delta/(2+\delta)}<\infty$, then $\sigma^2=E({Z_0}^2)+2\sum\limits_{j=1}^{\infty}E(Z_0Z_j)<\infty$, and if $\sigma\neq0$, then $\lim\limits_{n\to\infty}P\{\sigma^{-1}n^{-\frac{1}{2}}\sum\limits_{i=1}^{n}Z_i<z\}=\Phi(z)$.

Readers can be referred to Ibragimov (1971) for a proof and detailed discussion.

\bigskip
\noindent
{\large \bf A2: Proof of Theorem \ref{newBHWtest}}

\bigskip \noindent
Assume $\{x_{i,1},x_{i,2} , \cdots  ,x_{i,T}\}$ and $\{y_{j,1}, y_{j,2}, \cdots  ,y_{j,T}\},i\in \{1,2,\cdots,n_{1}\},j\in \{1,2,\cdots,n_{2}\}$ are both strong mixing stationary sequences whose
 mixing coefficient satisfying the conditions in Lemma 1. Then the following four sequences
\begin{eqnarray*}
&& \{Z_{1t} = n^{-1}\Big(\prod^{n_{1}}_{i=1} I \big(x_{i,t-L_{x_{i}}}^{m_{x_{i}}+L_{x_{i}}},x_{i,t+l-L_{x_{i}}}^{m_{x_{i}}+L_{x_{i}}},e  \big) \cdot
\prod^{n_{1}}_{j=1}I \big(y_{j,t-L_{y_{j}}}^{L_{y_{j}}},y_{j,t+l-L_{y_{j}}}^{L_{y_{j}}},e \big) - {C}_1 \big (M_{x}+{L_x},L_y,e;l \big )\Big)\} \, , \\
&& \{Z_{2t} = n^{-1}\Big( \prod^{n_{1}}_{i=1}I \big(x_{i,t-L_{x_{i}}}^{L_{x_{i}}},x_{i,t+l-L_{x_{i}}}^{L_{x_{i}}},e \big)
\cdot \prod^{n_{1}}_{j=1}I \big(y_{j,t-L_{y_{j}}}^{L_{y_{j}}},y_{j,t+l-L_{y_{j}}}^{L_{y_{j}}},e \big) - {C}_2  \big ({L_x},L_y,e;l \big )\Big)\} \, ,  \\
&& \{Z_{3t} = n^{-1}\Big( \prod^{n_{1}}_{i=1}I \big(x_{i,t-L_{x_{i}}}^{L_{x_{i}}},x_{i,t+l-L_{x_{i}}}^{L_{x_{i}}},e \big)
\cdot \prod^{n_{1}}_{j=1}I \big(y_{j,t-L_{y_{j}}}^{L_{y_{j}}},y_{j,t+l-L_{y_{j}}}^{L_{y_{j}}},e \big) - {C}_2  \big ({L_x},L_y,e;l \big ) \Big)\} \, ,  \\
&& \{Z_{4t} = n^{-1}\Big( \prod^{n_{1}}_{i=1} I \big(x_{i,t-L_{x_{i}}}^{L_{x_{i}}},x_{i,t+l-L_{x_{i}}}^{L_{x_{i}}},e \big) - {C}_4 \big ({L_x},e;l \big )\Big)\} \, , \\
&& t = L_{xy}+1, \cdots, T-l-L_{xy}-m+1,
\end{eqnarray*}
satisfy the conditions of Lemma 1, where $n=T-L_{xy}-l-m_{x}+1$ and
\begin{align*}
& {C}_1 (M_{x}+L_{x},L_{y},e;t,s)\equiv P\left(\parallel{ X}^{M_{x}+L_{x}}_{t-L_{x}}-{X}^{M_{x}+L_{x}}_{s-L_{x}}\parallel<e,\parallel{Y}^{L_{y}}_{t-L_{y}}-{ Y}^{L_{y}}_{s-L_{y}}\parallel<e\right) \, ,\\
& {C}_2 (L_{x},L_{y},e;t,s)\equiv P\left(\parallel{ X}^{L_{x}}_{t-L_{x}}-{ X}^{L_{x}}_{s-L_{x}}\parallel<e
,\parallel{ Y}^{L_{y}}_{t-L_{y}}-{ Y}^{L_{y}}_{s-L_{y}}\parallel<e\right)\\
& { C}_3 (M_{x}+L_{x},e;t,s)\equiv P\left(\parallel{ X}^{M_{x}+L_{x}}_{t-L_{x}}-{ X}^{M_{x}+L_{x}}_{s-L_{x}}\parallel<e\right)\\
& { C}_4 (L_{x},e;t,s) \equiv P\left(\parallel{ X}^{L_{x}}_{t-L_{x}}-{ X}^{L_{x}}_{s-L_{x}}\parallel<e\right) \, .
\end{align*}
So \{$Z_{1t}$\}, \{$Z_{2t}$\}, \{$Z_{3t}$\} and \{$Z_{4t}$\} satisfy the central limit theorem.

Further, for any real number $a_1, a_2, a_3$ and $a_4$, the sequence \{$Z_t = a_1 Z_{1t} + a_2 Z_{2t} + a_3 Z_{3t} + a_4 Z_{4t}, t = L_{xy}, \cdots, T-l-L_{xy}-m+1$\} also satisfies the conditions of Lemma 1 which implying that
\begin{eqnarray*}
 \sqrt{n}\left[
    \begin{array}{c}
\hat{C}_1  \big (m+{L_x},L_y,e;l \big )-{C}_1  \big (m+{L_x},L_y,e;l \big )\\
\hat{C}_2  \big ({L_x},L_y,e;l \big )-{C}_2  \big ({L_x},L_y,e;l \big )\\
\hat{C}_3  \big (m+{L_x},e;l \big )-{C}_3  \big (m+{L_x},e;l \big )\\
\hat{C}_4  \big (L_x,e;l \big )-{C}_4 \big ({L_x},e;l \big )
    \end{array} \right]
\overset{d}{\longrightarrow} N(0,\mathbf{\Sigma}),
\end{eqnarray*}
where
\begin{align*}
& {\hat{C}}_1 (M_{x}+L_{x},L_{y},e;l)\equiv \frac{1}{n}\sum^{T-l-m_{x}+1}_{t=L_{xy}+1}\prod^{n_{1}}_{i=1}I\left({x}^{m_{x_{i}}+L_{x_{i}}}_{i,t-L_{x_{i}}},{x}^{m_{x_{i}}+L_{x_{i}}}_{i,t+l-L_{x_{i}}},e\right)\cdot\prod^{ n_{2}}_{j=1}I\left({y}^{L_{y_{j}}}_{j,t-L_{y_{j}}},{y}^{L_{y_{j}}}_{j,t+l-L_{y_{j}}},e\right) \\
& {\hat{C}}_2 (L_{x},L_{y},e;l)\equiv \frac{1}{n}\sum^{T-l-m_{x}+1}_{t=L_{xy}+1}\prod^{n_{1}}_{i=1}I\left({ x}^{L_{x_{i}}}_{i,t-L_{x_{i}}},{x}^{L_{x_{i}}}_{i,t+l-L_{x_{i}}},e\right)\cdot\prod^{n_{2}}_{j=1}I\left({ y}^{L_{y_{j}}}_{j,t-L_{y_{j}}},{y}^{L_{y_{j}}}_{j,t+l-L_{y_{j}}},e\right)\\
& {\hat{C}}_3 (M_{x}+L_{y},e;l)\equiv \frac{1}{n}\sum^{T-l-m_{x}+1}_{t=L_{xy}+1}\prod^{n_{1}}_{i=1}I\left({ x}^{m_{x_{i}}+L_{x_{i}}}_{i,t-L_{x_{i}}},{x}^{m_{x_{i}}+L_{x_{i}}}_{i,t+l-L_{x_{i}}},e\right)\\
& {\hat{C}}_4 (L_{x},e;l)\equiv \frac{1}{n}\sum^{T-l-m_{x}+1}_{t=L_{xy}+1}\prod^{n_{1}}_{i=1}I\left({ x}^{L_{x_{i}}}_{i,t-L_{x_{i}}},{x}^{L_{x_{i}}}_{i,t+l-L_{x_{i}}},e\right)\\
\end{align*}
and $\mathbf{\Sigma}$ is a $4\times4$ symmetric matrix. Denote
$$ h_{1}(L_x,L_y,M_{x},l,k) = I(x^{L_x +M_{x}}_{L_{xy} +1+k-L_x}, x^{L_x +M_{x}}_{L_{xy}+1+k+l -L_x}, e) \, ,$$
$$ h_{2}(L_x,L_y,l,k) = I(y^{L_y}_{L_{xy}+1+k-L_y}, y^{L_y}_{L_{xy}+1+k+l -L_y}, e) \ ,$$
We have
\begin{eqnarray*}
&&\mathbf{\Sigma}_{11} = E \left[\big(h_{1}(L_x,L_y,M_{x},l,0)h_{2}(L_x,L_y,l,0) - {C}_1 (M_{x}+{L_x},L_y,e;l )\big)^2 \right]\\
&&+  \sum\limits_{k=1}^{n-1} 2(1-\frac{k}{n})E \Big[\big(h_{1}(L_x,L_y,M_{x},l,0)h_{2}(L_x,L_y,l,0) - {C}_1 (M_{x}+{L_x},L_y,e;l )\big)\\
&&\ \ \ \ \ \ \ \ \ \ \ \ \ \ \ \ \ \ \ \ \ \ \     \big(h_{1}(L_x,L_y,M_{x},l,k)h_{2}(L_x,L_y,l,k) - {C}_1 (M_{x}+{L_x},L_y,e;l )\big)\Big] \ ,
\\
&&\mathbf{\Sigma}_{12} = E \Big[\big(h_{1}(L_x,L_y,M_{x},l,0)h_{2}(L_x,L_y,l,0) - {C}_1 (M_{x}+{L_x},L_y,e;l )\big) \\
&& \ \ \ \ \ \ \ \ \ \ \ \ \ \big(h_{1}(L_x,L_y,0,l,0)h_{2}(L_x,L_y,l,0) - {C}_2 ({L_x},L_y,e;l )\big) \Big]\\
&&+  \sum\limits_{k=1}^{n-1} (1-\frac{k}{n})E \Big[\big(h_{1}(L_x,L_y,M_{x},l,0)h_{2}(L_x,L_y,l,0) - {C}_1 (M_{x}+{L_x},L_y,e;l )\big)\\
&&\ \ \ \ \ \ \ \ \ \ \ \ \ \ \ \ \ \ \ \ \ \  \big(h_{1}(L_x,L_y,0,l,k)h_{2}(L_x,L_y,l,k) - {C}_2 ({L_x},L_y,e;l )\big)\Big]\\
&&+  \sum\limits_{k=1}^{n-1} (1-\frac{k}{n})E \Big[\big(h_{1}(L_x,L_y,M_{x},l,k)h_{2}(L_x,L_y,l,k) - {C}_1 (M_{x}+{L_x},L_y,e;l )\big)\\
&&\ \ \ \ \ \ \ \ \ \ \ \ \ \ \ \ \ \ \ \ \ \  \big(h_{1}(L_x,L_y,0,l,0)h_{2}(L_x,L_y,l,0) - {C}_2 ({L_x},L_y,e;l )\big)\Big] \ ,
\\
&&\mathbf{\Sigma}_{13} = E \Big[\big(h_{1}(L_x,L_y,M_{x},l,0)h_{2}(L_x,L_y,l,0) - {C}_1 (M_{x}+{L_x},L_y,e;l )\big) \\
&& \ \ \ \ \ \ \ \ \ \ \ \ \ \big(h_{1}(L_x,L_y,M_{x},l,0) - {C}_3 ({M_{x}+L_x},e;l )\big) \Big]\\
&&+  \sum\limits_{k=1}^{n-1} (1-\frac{k}{n})E \Big[\big(h_{1}(L_x,L_y,M_{x},l,0)h_{2}(L_x,L_y,l,0) - {C}_1 (M_{x}+{L_x},L_y,e;l )\big)\\
&&\ \ \ \ \ \ \ \ \ \ \ \ \ \ \ \ \ \ \ \ \ \  \big(h_{1}(L_x,L_y,M_{x},l,k) - {C}_3 ({M_{x}+L_x},e;l )\big)\Big]\\
&&+  \sum\limits_{k=1}^{n-1} (1-\frac{k}{n})E \Big[\big(h_{1}(L_x,L_y,M_{x},l,k)h_{2}(L_x,L_y,l,k) - {C}_1 (M_{x}+{L_x},L_y,e;l )\big)\\
&&\ \ \ \ \ \ \ \ \ \ \ \ \ \ \ \ \ \ \ \ \ \  \big(h_{1}(L_x,L_y,M_{x},l,0) - {C}_3 ({M_{x}+L_x},e;l )\big)\Big] \ ,
\end{eqnarray*}
\begin{eqnarray*}
&&\mathbf{\Sigma}_{14} = E \Big[\big(h_{1}(L_x,L_y,M_{x},l,0)h_{2}(L_x,L_y,l,0) - {C}_1 (M_{x}+{L_x},L_y,e;l )\big) \\
&& \ \ \ \ \ \ \ \ \ \ \ \ \ \big(h_{1}(L_x,L_y,0,l,0) - {C}_4 ({L_x},e;l )\big) \Big]\\
&&+  \sum\limits_{k=1}^{n-1} (1-\frac{k}{n})E \Big[\big(h_{1}(L_x,L_y,M_{x},l,0)h_{2}(L_x,L_y,l,0) - {C}_1 (M_{x}+{L_x},L_y,e;l )\big)\\
&&\ \ \ \ \ \ \ \ \ \ \ \ \ \ \ \ \ \ \ \ \ \  \big(h_{1}(L_x,L_y,0,l,k) - {C}_4 ({L_x},e;l )\big)\Big]\\
&&+  \sum\limits_{k=1}^{n-1} (1-\frac{k}{n})E \Big[\big(h_{1}(L_x,L_y,M_{x},l,k)h_{2}(L_x,L_y,l,k) - {C}_1 (M_{x}+{L_x},L_y,e;l )\big)\\
&&\ \ \ \ \ \ \ \ \ \ \ \ \ \ \ \ \ \ \ \ \ \  \big(h_{1}(L_x,L_y,0,l,0) - {C}_4 ({L_x},e;l )\big)\Big] \ ,
\\
&&\mathbf{\Sigma}_{22} = E \left[\big(h_{1}(L_x,L_y,0,l,0)h_{2}(L_x,L_y,l,0) - {C}_2 ({L_x},L_y,e;l )\big)^2 \right]\\
&&+  \sum\limits_{k=1}^{n-1} 2(1-\frac{k}{n})E \Big[\big(h_{1}(L_x,L_y,0,l,0)h_{2}(L_x,L_y,l,0) - {C}_2 ({L_x},L_y,e;l )\big)\\
&&\ \ \ \ \ \ \ \ \ \ \ \ \ \ \ \ \ \ \ \ \ \ \     \big(h_{1}(L_x,L_y,0,l,k)h_{2}(L_x,L_y,l,k) - {C}_2 ({L_x},L_y,e;l )\big)\Big] \ ,
\\
&&\mathbf{\Sigma}_{23} = E \Big[\big(h_{1}(L_x,L_y,0,l,0)h_{2}(L_x,L_y,l,0) - {C}_2 ({L_x},L_y,e;l )\big) \\
&& \ \ \ \ \ \ \ \ \ \ \ \ \ \big(h_{1}(L_x,L_y,M_{x},l,0) - {C}_3 (M_{x}+{L_x},e;l )\big) \Big]\\
&&+  \sum\limits_{k=1}^{n-1} (1-\frac{k}{n})E \Big[\big(h_{1}(L_x,L_y,0,l,0)h_{2}(L_x,L_y,l,0) - {C}_2 ({L_x},L_y,e;l )\big)\\
&&\ \ \ \ \ \ \ \ \ \ \ \ \ \ \ \ \ \ \ \ \ \  \big(h_{1}(L_x,L_y,M_{x},l,k) - {C}_3 (M_{x}+{L_x},e;l )\big)\Big]\\
&&+  \sum\limits_{k=1}^{n-1} (1-\frac{k}{n})E \Big[\big(h_{1}(L_x,L_y,0,l,k)h_{2}(L_x,L_y,l,k) - {C}_2 ({L_x},L_y,e;l )\big)\\
&&\ \ \ \ \ \ \ \ \ \ \ \ \ \ \ \ \ \ \ \ \ \  \big(h_{1}(L_x,L_y,M_{x},l,0) - {C}_3 (M_{x}+{L_x},e;l )\big)\Big] \ ,
\end{eqnarray*}
\begin{eqnarray*}
&&\mathbf{\Sigma}_{24} = E \Big[\big(h_{1}(L_x,L_y,0,l,0)h_{2}(L_x,L_y,l,0) - {C}_2 ({L_x},L_y,e;l )\big) \\
&& \ \ \ \ \ \ \ \ \ \ \ \ \ \big(h_{1}(L_x,L_y,0,l,0) - {C}_4 ({L_x},e;l )\big) \Big]\\
&&+  \sum\limits_{k=1}^{n-1} (1-\frac{k}{n})E \Big[\big(h_{1}(L_x,L_y,0,l,0)h_{2}(L_x,L_y,l,0) - {C}_2 ({L_x},L_y,e;l )\big)\\
&&\ \ \ \ \ \ \ \ \ \ \ \ \ \ \ \ \ \ \ \ \ \  \big(h_{1}(L_x,L_y,0,l,k) - {C}_4 ({L_x},e;l )\big)\Big]\\
&&+  \sum\limits_{k=1}^{n-1} (1-\frac{k}{n})E \Big[\big(h_{1}(L_x,L_y,0,l,k)h_{2}(L_x,L_y,l,k) - {C}_2 ({L_x},L_y,e;l )\big)\\
&&\ \ \ \ \ \ \ \ \ \ \ \ \ \ \ \ \ \ \ \ \ \  \big(h_{1}(L_x,L_y,0,l,0) - {C}_4 ({L_x},e;l )\big)\Big] \ ,
\\
&&\mathbf{\Sigma}_{33} = E \left[\big(h_{1}(L_x,L_y,M_{x},l,0) - {C}_3 (M_{x}+{L_x},e;l )\big)^2 \right]\\
&&+  \sum\limits_{k=1}^{n-1} 2(1-\frac{k}{n})E \Big[\big(h_{1}(L_x,L_y,M_{x},l,0) - {C}_3 (M_{x}+{L_x},e;l )\big)\\
&&\ \ \ \ \ \ \ \ \ \ \ \ \ \ \ \ \ \ \ \ \ \ \     \big(h_{1}(L_x,L_y,M_{x},l,k) - {C}_3 (M_{x}+{L_x},e;l )\big)\Big] \ ,
\\
&&\mathbf{\Sigma}_{34} = E \Big[\big(h_{1}(L_x,L_y,M_{x},l,0) - {C}_3 (M_{x}+{L_x},e;l )\big) \\
&& \ \ \ \ \ \ \ \ \ \ \ \ \ \big(h_{1}(L_x,L_y,0,l,0) - {C}_4 ({L_x},e;l )\big) \Big]\\
&&+  \sum\limits_{k=1}^{n-1} (1-\frac{k}{n})E \Big[\big(h_{1}(L_x,L_y,M_{x},l,0) - {C}_3 (M_{x}+{L_x},e;l )\big)\\
&&\ \ \ \ \ \ \ \ \ \ \ \ \ \ \ \ \ \ \ \ \ \  \big(h_{1}(L_x,L_y,0,l,k) - {C}_4 ({L_x},e;l )\big)\Big]\\
&&+  \sum\limits_{k=1}^{n-1} (1-\frac{k}{n})E \Big[\big(h_{1}(L_x,L_y,M_{x},l,k) - {C}_3 (M_{x}+{L_x},e;l )\big)\\
&&\ \ \ \ \ \ \ \ \ \ \ \ \ \ \ \ \ \ \ \ \ \  \big(h_{1}(L_x,L_y,0,l,0) - {C}_4 ({L_x},e;l )\big)\Big] \ ,
\\
&&\mathbf{\Sigma}_{44} = E \left[\big(h_{1}(L_x,L_y,0,l,0) - {C}_4 ({L_x},e;l )\big)^2 \right]\\
&&+  \sum\limits_{k=1}^{n-1} 2(1-\frac{k}{n})E \Big[\big(h_{1}(L_x,L_y,0,l,0) - {C}_4 ({L_x},e;l )\big)\\
&&\ \ \ \ \ \ \ \ \ \ \ \ \ \ \ \ \ \ \ \ \ \ \     \big(h_{1}(L_x,L_y,0,l,k) - {C}_4 ({L_x},e;l )\big)\Big] \ .
\end{eqnarray*}

Under the null hypothesis, applying the delta method (Serfling, 1980), we have
\begin{align*}
\sqrt{n}\left(\frac{\hat{C}_1 \big (M_x+{L_x},L_y,e,l \big )}{\hat{C}_2 \big
({L_x},L_y,e,l \big )}-\frac{\hat{C}_3 \big (M_x+{L_x},e,l \big )}{\hat{C}_4 \big
({L_x},e,l \big )}\right) \overset{d}{\longrightarrow} N \big(0, \sigma^2
(M_x,{L_x},L_y,e,l) \big) \, ,
\end{align*}
where
$\sigma^2(M_{x},{L_x},L_y,e,l) = \nabla^{\prime}{\mathbf{\Sigma}}\nabla$, in which
\begin{align*}
\nabla &= \left( \frac 1 { C_2 \big({L_x},{L_y},e,l \big)} \, , \, - \frac
{ C_1 \big(M_{x}+{L_x},{L_y},e,l \big)}{C_2^2 \big({L_x},{L_y},e,l
\big)}  \, , \, - \frac 1 {C_4 \big({L_x},e,l \big)}  \, , \, \frac
{ C_3 \big({M_x}+{L_x},e,l \big)}{C_4^2 \big({L_x},e,l \big)} \right)^{\prime} \ .
\end{align*}
 An consistent estimator $\hat{\sigma}^2(M_{x},{L_x},L_{y},e,l)$ of the asymptotic variance
can be got by replacing all the parts in the sandwich $\nabla^{\prime}{\mathbf{\Sigma}}\nabla$ by their empirical estimates.

\end{document}